\documentclass[epsfig,amssymb,onecolumn]{aa}

\usepackage[dvips]{graphicx}

\def\EQ{\begin{equation}}
\def\EN{\end{equation}}
\def\EQA{\begin{eqnarray}}\def\ENA{\end{eqnarray}}

\title{An hydrodynamic shear instability in stratified disks}

\author{B. Dubrulle
\inst{1}
\and
L. Mari\'e \inst{1}
\and
Ch. Normand \inst{2}
\and D. Richard \inst{1,3} \and F. Hersant \inst{1,4}  \and J.-P. Zahn \inst{3}
}

\offprints{B. Dubrulle, bdubrulle@cea.fr}

\institute{GIT/SPEC/DRECAM/DSM, CNRS, URA 2464, CEA Saclay,
F-91191 Gif sur Yvette Cedex, France\\
\and
SPHT/DSM, CNRS, URA 2306 CEA Saclay, F-91191 Gif sur Yvette Cedex, France\\
\and
LUTH, Observatoire de Paris, F-92195 Meudon, France\\
\and
LESIA, Observatoire de Paris, F-92195 Meudon, France\\}

\authorrunning{Dubrulle et al.}
\titlerunning{Shear instability in stratified disks}

       \date{Received ; accepted }

\begin{document}

\abstract{
We discuss the possibility that astrophysical
accretion disks are
dynamically unstable to non-axisymmetric disturbances with
characteristic scales much smaller than the vertical scale height. The
instability is studied using three methods: one based on the energy
integral, which allows the determination of a sufficient condition of
stability, one using a WKB approach, which allows the determination
of the necessary and sufficient  condition for instability and a last one
by numerical solution. This linear instability  occurs in any inviscid
stably stratified
differential rotating fluid for rigid, stress-free or periodic 
boundary conditions, provided the
angular velocity $\Omega$ decreases outwards with radius $r$.
At not too small stratification, its
growth rate is a fraction of  $\Omega$. The influence of viscous 
dissipation and
thermal diffusivity on the instability is studied numerically, with emphasis on
the case when
$d \ln \Omega / d \ln r =-3/2$ (Keplerian case). Strong stratification
and large diffusivity are found to have a stabilizing effect.
The corresponding critical stratification and Reynolds number for the
onset of the instability in a typical disk are derived.
We  propose that the spontaneous generation of these linear modes is 
the source of turbulence in
disks, especially in weakly ionized disks.
}
\maketitle

\section{Introduction}

The simplest model of an accretion disk is that of a barotropic, axisymmetric
rotating shear flow in hydrostatic vertical equilibrium, with a
Keplerian velocity law. Realistic disks are also subject to
baroclinic effects (when the rotation departs from cylindrical, i.e.
when $\Omega$ varies
also with the axial coordinate $z$) and to a
vertical stratification, induced either by the hydrostatic state or
via the illumination of the surface of the disk due to the central
object. If the
stratification is unstable, it leads to turbulence via convective
instability. When the stratification is stable, it is generally
ignored or thought to be unimportant in the stability analysis, under
the rationale that it can only {\sl stabilize} the flow. Ignoring the
stratification and baroclinic effects makes the accretion disk look
like a simple
differentially rotating shear flow, with an azimuthal keplerian angular
velocity
profile $\Omega( r )\propto r^{-3/2}$ . Its linear stability with respect to
axisymmetric disturbances is governed by the Rayleigh criterion in the inviscid
limit:
\EQ
\frac{d(r^2 \Omega)^2}{dr}>0 \quad \hbox{for stability.}
\label{rayleigh}
\EN
Flows obeying this criterion are called {\sl
centrifugally stable}. Keplerian flow, in which angular momentum
increases outwards, falls into this category. Yet, there is
observational evidence that astrophysical (putatively Keplerian)
disks are turbulent ({\it cf.} Hersant et al. 2003), and thus that a source of
instability exists in
these flows.\

Leaving aside baroclinic effects, various mechanisms have been found
able to destabilize
centrifugally stable flows. They may or may not apply to
astrophysical disks.

\par i) Centrifugally stable flows can experience a globally subcritical
bifurcation (Dauchot \& Manneville 1997), induced by finite amplitude
disturbances involving non-linear mechanisms not captured by the Rayleigh
criterion (Dubrulle 1993). The transition threshold in this case is
related to the amplitude
of the external disturbance, as typically observed in a plane shear
flow~ (Dauchot \& Daviaud 1994 ).

Taylor-Couette experiments, with fluid sheared between two concentric
rotating cylinders in the centrifugally stable regime, have revealed such
  transition. When the inner cylinder is at rest, the data of
Wendt (1933) and Taylor (1936), re-analyzed by Richard and
Zahn (1999), show that for the typical amplitude of the
intrinsic disturbances of the experimental devices, turbulence subcritically
sets in as soon as
$ R=U d/\nu > R^{nl}_c $, where $U$ is the relative velocity between the
walls, $d$, the radial extent of the flow (the gap) and $\nu$ the viscosity.

For small curvature, the threshold is essentially independent of the
gap/radius ratio and in the limit of plane Couette flow, $R^{nl}_c \simeq
1500$, a value in agreement with the minimal Reynolds number for which
turbulence can be induced in plane Couette flow. For large curvature, $
R^{nl}_c $ scales with the square of the gap/radius ratio.

Finally, novel laboratory experiments were performed
recently (Richard 2001), to refine the present analysis and to
explore regimes with co-rotating cylinders. For the first time, the
hysteretic behavior of the transition with the inner cylinder at rest has been
described, a signature of subcriticality and turbulent regimes were detected
when the angular velocity decreases outward, in the centrifugally stable
region.\

ii) Centrifugally stable flows can also be destabilized by
compressibility effects via non-axisymmetric instabilities
(Papaloizou \& Pringle 1984). This instability occurs independently
of boundary conditions as long as $d\ln\Omega/d\ln r<-\sqrt{3}$
(Papaloizou \& Pringle, 1985; Glatzel, 1987). However, the existence
of such an instability for Keplerian disks requires the presence of
at least one sharp edge (Goldreich \& Narayan, 1985). This condition
may be unrealistic in standard accretion disks.\

\begin{figure}[hhh]
\centering
\includegraphics[width=8.5cm]{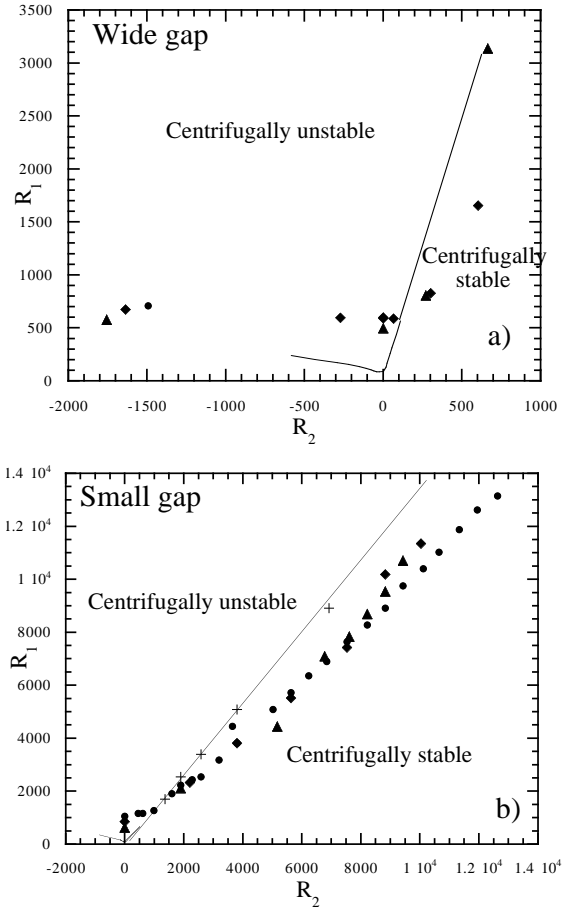}
\caption[]{Neutral-stability curves in  a stratified Taylor-Couette flow
experiment for different stratifications, in the domain $(R_2,R_1)$
where $R_2$ and $R_1$ are the
Reynolds numbers based on the gap size and the velocity at the outer
or inner cylinder. a) In the wide gap limit (ratio of inner cylinder
to outer cylinder equal to $0.2$), experiment
of Whithjack and Chen (1974): $N=1.5
s^{-1}$ (circles), $N=1.25 s^{-1}$ (diamonds) and $N=0.88s^{-1}$
(triangles); b) in the small gap limit (ratio of inner cylinder to
outer cylinder equal to $0.8$), experiment of  Boubnov and Hopfinger
(1995) $N=1.21
s^{-1}$ (circles), $N=0.89 s^{-1}$ (diamonds),  $N=0.54 s^{-1}$
(triangles) and $N=0 s^{-1}$ (crosses).
The continuous line is the neutral-stability line for centrifugal
instability  computed by Snyder
(1968) in an un-stratified Taylor-Couette experiment.
      In the centrifugally unstable case, one sees that the stratification
enhances stability. In the centrifugally stable case, stratification
favors instability.}
   \label{fig:figure1}
\end{figure}

\par iii) Centrifugally stable flows can be further destabilized by
adding another restoring force, which acts as a  {\sl catalyzer}.
      A first example is a vertical
magnetic field (Velikhov 1959; Chandrasekhar 1960),
which renders such flows unstable provided they are anticyclonic,
i.e. if $d (\Omega)^2 /dr <0$. The
application of this mechanism to disks was first discussed by
Balbus and Hawley (1991); they showed that  the stratification of the disk does
not modify the result, and  that the maximal growth rate of
instability in that case is of the order of the angular
rotation velocity in the inviscid limit.
The instability of a rotating flow subject to vertical
magnetic field includes a surprising paradox: experimentally, it was
found that in liquid metals, in the centrifugally
unstable case, the magnetic field inhibits
instabilities, and thus, has a {\sl stabilizing} influence
(Chandrasekhar 1960). In the
centrifugally stable case, it creates an instability. Summarizing, it
appears that a stabilizing factor acts upon a stable
flow so as to generate instability.

iv) A similar behavior was observed in rotating flows subject to a
vertical stable stratification (Whithjack \& Chen 1974, Boubnov and
Hopfinger, 1995). In the
{\sl centrifugally unstable} case, the stratification enhances the
stability of the flow and tends to increase the critical Reynolds
number of the inner cylinder (Fig. \ref{fig:figure1}). In the
{\sl centrifugally
stable} regime, the experimental stability curve crosses the critical
line for the Rayleigh criterion, and turbulence sets in
via a non-axisymmetric mechanism. This instability is present in the
small gap and wide gap regime, showing that curvature effect do not
play a role in the instability mechanism. A possible theoretical
explanation for this instability has been given recently by Molemaker et al.
(2001) and Yavneh et al (2001). They performed an analytical and
numerical stability analysis of a rotating flow in the presence of a
stable vertical temperature gradient, and discovered the existence of a linear
non-axisymmetric instability for all
anticyclonically sheared flows. In their work, they interpret this instability
as being caused by the interaction of two edge modes through a
mechanism of arrest and phase locking along the boundaries by the
mean shear flow.\

We note that in astrophysical context, the stability of plane
rotating shear flow has often been tackled through the so-called
shearing sheet transformation introduced by Goldreich and Lynden-Bell
(1965). In this approximation, the disk structure is equivalent to a
rotating plane Couette flow, with shear being $-3/2$ of the
rotation. The advantage of this kind of analysis is the possibility
to investigate stability, by using a
special class of disturbances for which the stability analysis of
this flow can be reduced to study of ordinary differential equations.
  These modes, first introduced by Lord Kelvin in 1887, have the shape
$u(x,y,z,t,k,\beta,k_z)=u(t)\exp(i(k-S\beta t)x+i\beta y+ik_z z)$.
Here $S$ is the Couette shear rate, with the flow velocity in the 
$y$-direction varying with $x$.
Individual non-axisymmetric
such modes have been found to exhibit algebraic transient growth
under various conditions, such as compressibility (Dubrulle \&
Knobloch 1992) or vertical stable and unstable stratification
(Knobloch 1984; Korycansky 1992). However, any amount of
dissipation causes ultimate decay of these modes, due to shear
winding of the azimuthal wave-numbers. From a theoretical point of
view, it can be shown that these modes are associated with the
non-normality of the operator governing the perturbation dynamics.
They may be the basis of a noise amplification mechanism, resulting
in significant angular momentum flux (Ioannou \& Kakouris 2001).
However, these modes do not form a complete basis (for example, they
cannot describe perturbations with periodic boundary conditions in
$x$), and their study alone is not sufficient to determine the
stability properties of the flow, as recognized by Korycansky (1992).
A complete stability analysis requires solving the ``initial
condition" problem, looking at the stability properties of an infinite
superposition of sheared modes, moving with the flow. However, even in
the case of non-stratified rotating shear flow, this problem is
difficult to handle analytically and calls for a numerical solution
(Cambon et al. 1994). Therefore, the
shearing-sheet/sheared mode method provides only a partial answer to
the stability problem.\

In the present paper, we resort to other approaches, in which
boundary conditions must be fixed a priori, but which allow for a
complete treatment of the instability. Our approach
is complementary to the usual sheared mode approximation.
We use two classical, analytically tractable approaches
to reexamine the instability of a stratified astrophysical
disk: one based on the energy method, the second based on the WKB
approximation. The energy method is valid for a wide class of
boundary conditions, namely rigid (vanishing velocity at the
domain boundary), stress free (vanishing velocity derivative at
the domain boundary) or periodic boundary conditions in the shear
direction. Therefore, it does not apply to individual sheared modes,
which satisfy none of these requirements.  This method leads to a
sufficient condition for stability valid in the inviscid limit, and
for perturbations with a characteristic scale much smaller than the
vertical scale height (Section 2.1). The influence of curvature on
this instability is discussed in
Section 2.2. In Section 3, we use a method based on the WKB approximation
to derive an explicit solution of the stability problem and exhibit
unstable modes in the parameter space covering the sufficient
condition for stability. This shows that the sufficient condition for
stability is probably necessary.
This theoretical study is completed in section 4 by a numerical study,
probing instability regimes and the influence of dissipative
processes on the instability. This study
enlarges the numerical study of Yavneh et al. (2001) towards conditions
more typical of astrophysical disks, namely Keplerian velocity
profile and finite Prandtl number.
In Section 5, we discuss the importance of this mechanism
for turbulence generation in disks, and
the similarity between this instability and the instability generated
by a vertical magnetic field. Our conclusions follow in Section 6,
where the interplay between the present shear instability and its
baroclinic counterpart is discussed.

\section{Theoretical study}

We consider a differentially rotating compressible stratified disk.
In the following, we focus on perturbations with typical radial scales
small compared with the vertical scale height of the disk. This limit
provides two simplifications: first, it allows us to consider only
the barotropic case, in which all vertical  dependence of the
equilibrium quantities is ignored. Then it allows elimination of
acoustic waves and we can
work in the Boussinesq approximation (Korycansky 1992). The
dynamical equations ruling the perturbation ${\bf u }$
take then the simple form
\EQA
\nabla\cdot {\bf u}&=&0,\nonumber\\
D_t {\bf u} +{\bf u}\cdot \nabla {\bf U}+
2 \Omega {\bf e}_z \times  {\bf u} + \nabla p
- h {\bf g} &=&\nu \Delta {\bf u},\nonumber\\
D_t h +{\bf u}\cdot\nabla H&=&\frac{\nu}{Pr}\Delta h,
\label{basicequation}
\ENA
in a frame rotating with the constant angular velocity $\Omega$ (to
be chosen later).
Here $D_t=\partial_t +{\bf U} \cdot \nabla$ is the total derivative,
$\nu$ is the viscosity, $Pr$ is the Prandtl number,
$ {\bf g} = \nabla P /\rho$ the local effective gravity,
$H$ the basic stratification, and $h$ is
the stratification perturbation in the Boussinesq approximation.\

The instability we are interested in is present both in rotating
Couette flow (Kushner et al. 1998), i.e. for plane geometry, or in
Taylor-Couette
flow (Molemaker et al. 2001, Yavneh et al. 2001) i.e. for circular
geometry. This suggests that curvature effects do not play a role in
the instability. To clarify the presentation, we first
deal with the simpler, plane case in Section 2.1. In the
astrophysical context, this case is equivalent to the so-called
shearing sheet approximation of Goldreich and Lynden-Bell (1972). It
is also
the small gap limit of the Taylor-Couette flow, and is relevant to
many laboratory experiments. After discussion of this simple,
illustrative case, we come back to the large gap limit (relevant to
disks) in Section 2.2, where curvature effects are included. For
simplicity, we also consider in this Section only the inviscid limit
$\nu=0$, keeping the general case
of finite viscosity for numerical exploration (Section 3).

\subsection{The plane Couette case}

In this case, we assume Cartesian geometry $x,y,z$. The basic flow is
a combination of a pure shear flow
\EQ
{\bf U}=Sx e_y,
\label{localapprox2}
\EN
with a constant rotation $\Omega$ along the $z$ axis.\

We expand the perturbation into normal modes:
$({\bf u},h) =({\bf \hat u},\hat h)(x)\,
\exp[ i (\beta y+k_z z-\omega t)]$. Plugging this into
(\ref{basicequation}) and noting the
      velocity perturbation $\hat {\bf u} = (u, v, w)$ and the rescaled
      entropy perturbation $\gamma \hat h= \theta\partial_z H$, we obtain
the following system:
\EQA
-i\sigma u-2\Omega v&=&-Dp \nonumber\\
-i\sigma v+Z u&=&-i\beta p\nonumber\\
-i\sigma w +N^2 \theta&=& -i k_z p,\nonumber\\
-i\sigma \theta&=& w,\nonumber\\
Du+i\beta v+i k_z w&=&0.
\label{basicequationlocal}
\ENA
Here, we have introduced the $x$-derivative $D$ and the Doppler-shifted
frequency $\sigma=\omega-\beta Sx$. The equation also involves the
frequency $Z=2\Omega+S$ and the squared Brunt-V\"ais\"al\"a frequency in
the axial direction:
\EQ
N^2=- {1\over \gamma} g_z \, \partial_z H,
\label{Bvdefinition}
\EN
which is positive in the case of stable stratification. The system of
equations (\ref{basicequationlocal}) also describes a Keplerian disk
in the shearing sheet approximation, provided $\Omega$ is the local
Keplerian rotation and $Z=\Omega/2$.\

The system (\ref{basicequationlocal}) is a coupled differential
system. To investigate the basic mechanism of instability, we use
(\ref{basicequationlocal}) to eliminate three components in order to
get a single differential system for one component, say $u$.
For this, we first eliminate the pressure by using the three last
equations of (\ref{basicequationlocal}) to get:
\EQ
p=-i\frac{N^2-\sigma^2}{k_z^2\sigma}\left(Du+i\beta v\right).
\label{pressure}
\EN
We then substitute this expression into the second equation of
(\ref{basicequationlocal}) to obtain $v$ as a function of $u$. This
expression can be used in the first equation of
(\ref{basicequationlocal}) to obtain a closed equation for $u$. After
simplification and rearrangements, we obtain finally:
\EQ
D\left(\frac{K}{G}Du\right)+\frac{\beta Z}{\sigma}uD\left(\frac{K}{G}\right)+
2k_z^2 \frac{\Omega Z}{G} u-u=0,
\label{finalu}
\EN
where $K=\sigma^2-N^2$ and $G=(k_z^2+\beta^2)\sigma^2-\beta^2 N^2$.
It is easy to check that this equation is the zero-curvature limit of
that derived by Yavneh et al. (2001).
Note that both $K$ and $G$ depend on $x$ through $\sigma$.\

Eq. (\ref{finalu}) is a classical generalized eigen-value problem
(corresponding to a linear differential operator) which can be solved
once the boundary conditions have been specified. The instability
occurs for values of the eigen-values such that $\omega$, and
therefore $\sigma$ has a positive imaginary part.\

To derive stability conditions, we use an integral method adapted
from Chandrasekhar
(1960).
We first reset equation (\ref{finalu}) into a standard shape by the
change of function $u=\hat u \left(\frac{K}{G}\right)^{-1/2}$ to
obtain:
\EQA
&D^2& \hat u-E\hat u=0,\nonumber\\
&E&=\frac{a^2}{4}+\frac{Da}{2}-\frac{a\beta Z}{\sigma}-
2k_z^2 \frac{\Omega Z}{K} +\frac{G}{K}
\label{finaluhat}
\ENA
where $a=D(\ln(K/G)$. We then multiply (\ref{finaluhat}) by the
complex conjugate $\hat u^*$ and integrate over the whole domain. We
obtain an integral equation, valid for both rigid ($\hat u=0$ at
boundaries), free-slip ($Du=0$ at boundaries) or periodic boundary
conditions:
\EQ
\int dx \left | D \hat u\right|^2+\int dx \, E \left|\hat u\right |^2= 0.
\label{integratecondition}
\EN
Both $\left|D\hat u\right|^2$, and $\left|\hat u\right|^2$ are
positive continuous real function over the integration domain. The
function $E$ is a complex function of $x$ (through $\sigma$) whose
real and imaginary part $E_r$ and $E_i$ are continuous over the
integration domain. By continuity (intermediate value theorem), there
exist two points $x_1$ and $x_2$ such that:
\EQA
&&\int dx \left | D \hat u\right|^2+E_r(x_1)\int dx  \left |\hat
u\right |^2\equiv
I_1+E_r(x_1)\, I_0= 0,\nonumber\\
&&E_i(x_2)\int dx  \left |\hat u\right |^2= 0.
\label{intermediate}
\ENA
In principle, a stability condition may then
be found by imposing that there exist no solutions with positive $\omega_i$
for any value of $x_2$ and $x_1$. In practise, the algebraic
equation is of order eight in $\omega_i$, and we were not able to
derive
simple explicit stability conditions. We therefore restrict ourselves
to the instability condition in the limit of weak non-axisymmetry, by
using the fact that $\omega_i$ vanishes in the limit $\beta=0$ and
that the instability is non-oscillatory $\omega_r=0$ (see Molemaker
et al, 2001). Then,
since $\sigma=-\beta S x+i\omega_i$, we can set $\sigma=\beta
\tilde \sigma$ and expand $E$ as a function of $\beta$. To first
order in $\beta$, we find:
\EQ
\frac{E}{k_z^2}=2\frac{\Omega Z}{N^2}-4\frac{S\Omega}{k_z^2\tilde \sigma^2-N^2}
+3 \left(\frac{S N}{k_z^2\tilde \sigma^2-N^2}\right)^2.
\label{expanE}
\EN
To obtain an analytical condition, we
set $\rho e^{i\psi}=k_z^2\tilde \sigma^2-N^2$ and note that finding
positive $\omega_i$ as a function of $x_2-x_1$ is equivalent to finding
$\rho$ as a function of $\psi$. In these variables, the first
equation (\ref{intermediate}) becomes to first order in $\beta$:
\EQ
\left(\frac{I_1}{k_z^2I_0}\frac{N^2}{4\Omega^2}+\frac{S}{2\Omega}\right)
\rho^2+\left(\rho-
\frac{SN^2}{2\Omega}\cos\psi\right)^2+\left(\frac{S}{2\Omega}\right)^2
(5\cos^2\psi-3)N^4=0.
\label{seconddegrejolie}
\EN
In the limit of vanishing stratification, the last term of the l.h.s.
of eq. (\ref{seconddegrejolie}) is negligible. The second term of the
l.h.s. is always positive. So a (sufficient) condition which guarantees
the absence of unstable solutions is that the first term of the
l.h.s. is positive, i.e.:
\EQ
\frac{S}{2\Omega}>-
\frac{1}{4}\frac{I_1}{k_z^2I_0}\frac{N^2}{\Omega^2}\quad\rm{STABILITY}
.
\label{condiinstability}
\EN
Using the energy method, we have derived a sufficient criterion for
stability for periodic, stress-free or rigid boundary conditions.
Therefore, it does not prove that flows with  $S/2\Omega<0$ are
unstable. However, in the WKB approximation of Section 3 and
numerical stability analysis performed in Section 4, we found that
all flows with $S/2\Omega<0$ are unstable. This shows that condition
(\ref{condiinstability}) is also a necessary condition for stability
for these boundary conditions.\

In the limit of vanishing stratification, we find $S/2\Omega>0$ as
a criterion for stability, instead of the $S/2\Omega>-1$ criterion
for centrifugal stability. The stratification thus enlarges the
domain of instability. However, this criterion is
non-axisymmetric;
for $\beta=0$, the only non-trivial modes are stable, with
the epicyclic frequency $\sigma^2 =( 2 \Omega Z) k_z^2$.
Note that our criterion is in agreement with
the criterion derived by Molemaker et al (2001) and Yavneh et al
(2002), which was $S/2\Omega<0$ for instability.
Note also that as the
stratification increases, it becomes increasingly easy to
satisfy (\ref{condiinstability}). So, while a weak stratification
destabilizes the flow, a large stratification re-stabilizes it. We
show in Section 2.6 that when the stratification is replaced by a
vertical magnetic field, a similar phenomenon occurs: weak fields are
destabilizing, while very large fields are stabilizing.

\subsection{Influence of curvature}
We now explore how curvature effects modify the instability. For
this, we assume cylindrical coordinates $r,\phi,z$ and consider the
stability of a general rotating shear flow
\EQ
{\bf U}=r\Omega(r) e_\phi,
\label{totalflow}
\EN
Thus, the rotation frequency is noted $\Omega$, and the shear rate
$S=r\partial_r \Omega$.
The normal-mode decomposition is
\EQ
({\bf u},h) =({\bf \hat u},\hat h)(x)
e^{ i (l\phi+k_zz-\omega t)}.
\label{normodes}
\EN
The analog of (\ref{finalu}) in this general case is derived in
Yavneh et al 2001. It is:
\EQA
&&D\left(\frac{K}{G} rD\left(ru\right)\right)\nonumber\\
&&+\left(\frac{l}{\sigma}D\left(\frac{KZ}{G}\right)+
2rk_z^2 \frac{\Omega Z}{G} -\frac{1}{r}\right)(ru)=0,
\label{finaluc}
\ENA
where $Z=2\Omega+S$, $K=\sigma^2-N^2$, $G=(r^2k_z^2+l^2)\sigma^2-l^2
N^2$, $\sigma=\omega-l\Omega$ and $D$ is the derivative with respect
to $r$.\

Following a method proposed by Dubrulle and Graner (1994), we now
perform the change of variable $x=\ln(r/r_0)$, where $r_0$ is any
characteristic radius, and the change of function $\tilde u= ru$.
Setting $\tilde k_z= k_z r_0 \exp{2x}$, we may then write (\ref{finaluc})
as:
\EQ
D_x\left(\frac{K}{G}D_x\tilde u\right)+\frac{l }{\sigma}\tilde
uD\left(\frac{KZ}{G}\right)+
2\tilde k_z^2 \frac{\Omega Z}{G} \tilde u-\tilde u=0,
\label{finalutilde}
\EN
similar to (\ref{finalu}), the equation for the
plane case. The main difference here is that $\tilde k_z$, $Z$,
$\Omega$ and $S$ are now functions of $x$. Keeping this in mind, we
then repeat the same steps as in Section 2.1
until we get the equivalent of (\ref{intermediate}) with the first
order expansion of E given by:
\EQA
E&=&1+2\tilde k_z^2\frac{\Omega Z}{N^2}-\frac{DZ}{\tilde\sigma}\nonumber\\
&+&\frac{N^2+\tilde k_z^2\left(\tilde\sigma DS-2\tilde\sigma
S-4S\Omega\right)}{\tilde k_z^2\tilde \sigma^2-N^2}
+3 \tilde k_z^2 \left(\frac{(\tilde\sigma-S) N}{\tilde k_z^2\tilde
\sigma^2-N^2}\right)^2.
\label{expanEc}
\ENA
    From (\ref{expanEc}), we see that the only effects of curvature occur
at small $\tilde k_z$. Indeed, for large $\tilde k_z$,
$\tilde\sigma\tilde k_z=O(1)$ , we can expand
(\ref{expanEc}) and get to first order in $\tilde k_z$:
\EQ
E=2k_z^2\frac{\Omega Z}{N^2}-4k_z^2\frac{S\Omega}{N^2(k_z^2\tilde
\sigma_0^2-1)}
+3 k_z^2 \left(\frac{S N}{(k_z^2\tilde \sigma_0^2-1)N^2}\right)^2,
\label{expanEfin}
\EN
which exactly matches the expression obtained in the plane Couette
(\ref{expanE}). We thus conclude that a condition for stability at
large $\tilde k_z$ is
(\ref{condiinstability}). Note that in astrophysical thin disks
$\tilde k_z\sim (R/H)\exp{x}$, and so the curvature effects are
likely to be small everywhere except in the innermost part of the
disk. In that region, however, many other physical processes may play
an important role (like stellar magnetic field, general relativity
effects, radiative processes,..) so that we do not think it
very important to explore this peculiar effect.

\section{WKB approximation}
\subsection{Method}
In the previous section, we derived stability conditions independent
of the explicit shape of the solutions. In this Section, we derive analytical
solutions using a WKB approximation (Bender \& Orszag 1987, Nayfeh
1978). This allows us to exhibit explicit examples of unstable cases with
$S/2\Omega<0$. The WKB approximation requires a small parameter. In
the following, it will be convenient to choose
$\varepsilon=S/\Omega$ and let $\varepsilon \to 0$.
We introduce the
slow horizontal variable $X=\varepsilon x$.
In the spirit of WKB approximations, the solutions for $u$ and $p$
will be split into a fast varying complex phase 
$\Theta(X)/\varepsilon$ and slowly
varying functions $U(X)$
and $P(X)$ according to
\begin{equation}
(u(x), p(x)) = (U(X), P(X)) \exp(i\Theta(X)/\varepsilon) \label{wkb}
\end{equation}
As a result, the derivative of $u$ and $p$ are replaced by
$Du=i\Theta_X U +\varepsilon U_X$ and
$Dp=i\Theta_X P +\varepsilon  P_X$ (with $f_X=\partial_X f$. This
development must be plugged into the equations of motion
(\ref{basicequationlocal}). For this, we rewrite them as a second
order differential system for the velocity
component $u$ and the pressure $p$ alone
\begin{eqnarray}
(2 \Omega Z -\sigma^2) u &=& -2i \beta \Omega p + i \sigma Dp \label{Dp}\\
(N^2 - \sigma^2)( \sigma Du +\beta Z u)&=& i \left[ \sigma^2 k_z^2
-\beta^2(N^2 - \sigma^2)\right]p \label{Du}
\end{eqnarray}
For non vanishing azimuthal wavenumbers $\beta \ne 0$ the quantity
$\sigma$ depends on
the $x$-coordinate. We take
advantage of the particular scaling for the growth rate: 
$\omega=\beta \Omega \bar \omega$
leading to $\sigma=\beta \Omega \bar \sigma$ with $\bar \sigma = \bar 
\omega- \varepsilon x$. After insertion
into eq. (\ref{Dp})-(\ref{Du}), we obtain:
\begin{eqnarray}
\Omega(2 (2+\varepsilon) -\beta^2 \bar \sigma^2) U &=& -2i \beta P +
\beta \bar \sigma (i \varepsilon P_X -\Theta_X P) \label{PX}\\
\Omega(1 - Fr^2 \beta^2\bar \sigma^2)(\varepsilon \bar \sigma U_X
+(2+\varepsilon +i \bar \sigma \Theta_X)U)
&=& i \beta \bar G P ,  \label{UX}
\end{eqnarray}
where $\bar G= \alpha^2 Fr^2 \bar \sigma^2 - 1$ with
$\alpha^2=k_z^2+\beta^2$, and the Froude number being $Fr = \Omega / 
N$. The slowly varying
functions $U(X), P(X)$ are then expanded as
\begin{equation}
U = U_0 +\varepsilon U_1 +... \quad \mbox{and} \quad
P = P_0 +\varepsilon P_1 +... \label{expan}
\end{equation}
Upon substituting (\ref{expan}) into the governing Eqs.
(\ref{PX})-(\ref{UX}) one is led to
a set of successive problems for $(U_0,P_0)$, $(U_1,P_1)$ etc... .

At the leading order, the governing equations for $U_0,P_0$ are
\begin{eqnarray}
\Omega(4 -\beta^2 \bar \sigma^2)U_0 &=& -\beta (2i+\bar \sigma
\Theta_X)P_0 \label{U0}\\
\Omega(1 - Fr^2\beta^2 \bar \sigma^2)(2i-\bar \sigma \Theta_X)U_0&=&-
\beta \bar G P_0 \label{P0}
\end{eqnarray}
The above system admits non-trivial solutions provided that $\Theta_X$
is given by
\begin{equation}
\Theta_X^2 = -{\left(4 k_z^2 Fr^2 +\beta^2 -\bar \sigma^2 Fr^2 \alpha^2
\beta^2\right) \over
(1 - \beta^2 Fr^2 \bar \sigma^2)} \label{TX2}
\end{equation}
Once $\Theta$ is known the slowly varying functions $U_0(X)$ and
$P_0(X)$ are determined by the
condition that there are non-trivial solutions for $U_1(X)$ and $P_1(X)$.
At the order $\varepsilon$ the governing equations are
\begin{eqnarray}
\Omega(4 -\beta^2 \bar \sigma^2)U_1  + \beta (2i+\bar \sigma \Theta_X)P_1 &=&
i \beta \bar \sigma P_{0X}-2\Omega U_0\\
\Omega(1 - Fr^2\beta^2 \bar \sigma^2)(2i-\bar \sigma \Theta_X)U_1+
\beta \bar G P_1&=&
-i(1 - Fr^2\beta^2 \bar \sigma^2)(\bar \sigma U_{0X} +U_0)
\end{eqnarray}
The existence condition is thus
\begin{equation}
(i \beta \bar \sigma P_{0X}-2\Omega U_0)\bar G +i(1 - Fr^2\beta^2
\bar \sigma^2)
(\bar \sigma U_{0X} +U_0)(2i+\bar \sigma \Theta_X)=0 \label{exist}
\end{equation}
The following identities
\begin{equation}
P_0=-\Omega{(4 -\beta^2 \bar \sigma^2) \over \beta (2i+\bar \sigma
\Theta_X)} U_0 \equiv -H U_0
\quad \mbox{and}
\quad {\bar G \over (1 - Fr^2\beta^2 \bar \sigma^2)}\equiv{(2i-\bar
\sigma \Theta_X) \over H}
\end{equation}
are used to transform (\ref{exist}) in an equation for $U_0$ alone
\begin{equation}
2{U_{0X} \over U_0} +{H_{X} \over H}+ {1 \over \sigma}+
{2i(\bar \sigma \Theta_X)_X \over \bar \sigma \Theta_X (2i+\bar
\sigma \Theta_X)}+
{2 i \over \Theta_X}\left[{k_z^2 Fr^2 \over (1 - Fr^2 \beta^2 \bar \sigma^2)}
   -{2\beta^2 \over (4 -\beta^2 \bar \sigma^2)}\right]=0 \label{exu0}
\end{equation}
Equation (\ref{exu0}) is a closed differential equation providing
the analytical shape of the modes. Solutions corresponding to $\bar
\sigma$ having a positive imaginary part provide the unstable modes.
This requires the a priori knowledge of $\Theta(X)$. This expression
is independent of $X$ in a few cases discussed below. These cases
provide analytical expression of unstable modes, as we now show.

\subsection{Exponential solutions}
In this Section, we investigate the exponential solution, obtained when
$\Theta$ is purely imaginary. This case is readily obtained in at
least two cases.
For the particular
value of the Froude number $Fr=1/2$, one may check that $\Theta_X^2=-\alpha^2$.
Also, in the limit: $\beta \to 0$ and $Fr \approx {\cal O} (1)$,
considered previously by
Yavneh et al. one gets $\Theta_X^2=-4 k_z^2 Fr^2$. These two cases
being formally identical, we focus on the last one, which has been
intensively studied by Yavneh et al.
Taking into account the constancy of  $\Theta_X$,
the equation (\ref{exu0})is readily integrated to give the result
\begin{equation}
   U_0=C(1\pm k_z Fr \bar \sigma) \exp(\mp k_z Fr X)
\end{equation}
where $C$ is a constant of integration. Upon substituting the above 
expression in (\ref{wkb})
one recovers the edge modes of Yavneh et al.
\begin{equation}
U=(1\pm k_z Fr \bar \sigma) \exp[\mp 2 k_z Fr (1 + \varepsilon /2)x]
\end{equation}
written here in the asymptotic limit $(1 + \varepsilon)^{1/2} \to 1 +
\varepsilon /2$. From their work, we thus get that unstable solutions
exist for all anticyclonic flows $S/2\Omega<0$, while no unstable
solution exist in this limit for $S/2\Omega\ge 0$. This shows that
the criterion for stability derived in Section 2 is also necessary in
this limit.

\subsection{Oscillating solutions}

The oscillatory situation will correspond to $\Theta$ real values
leading to solutions in (\ref{wkb}) which are oscillating in space.
   They occur for example when $\beta$ is large while $Fr \to 0$
and $\beta Fr$ is of order unity so expression (\ref{TX2}) is
approximated by
\begin{equation}
\Theta_X^2 = -\beta^2 \left(1 -  {k_z^2 Fr^2 \bar \sigma^2  \over
(1 - \beta^2 Fr^2 \bar \sigma^2)}\right) \label{Tb}
\end{equation}
According to the prescribed limit the second term in the r.h.s. of
(\ref{Tb}) has a small
denominator and a numerator of order unity so it is dominant and
$\Theta_X$ can be written
\begin{equation}
\Theta_X = \pm {\beta k_z Fr \bar \sigma  \over (1 - \beta^2 Fr^2
\bar \sigma^2)^{1/2}} , \label{TXa}
\end{equation}
and after integration one gets
\begin{equation}
\Theta = \pm { k_z \over  \beta Fr}  (1 - \beta^2 Fr^2 \bar \sigma^2)^{1/2} .
\end{equation}

\begin{figure}[hhh]
\centering
\includegraphics[width=8.5cm]{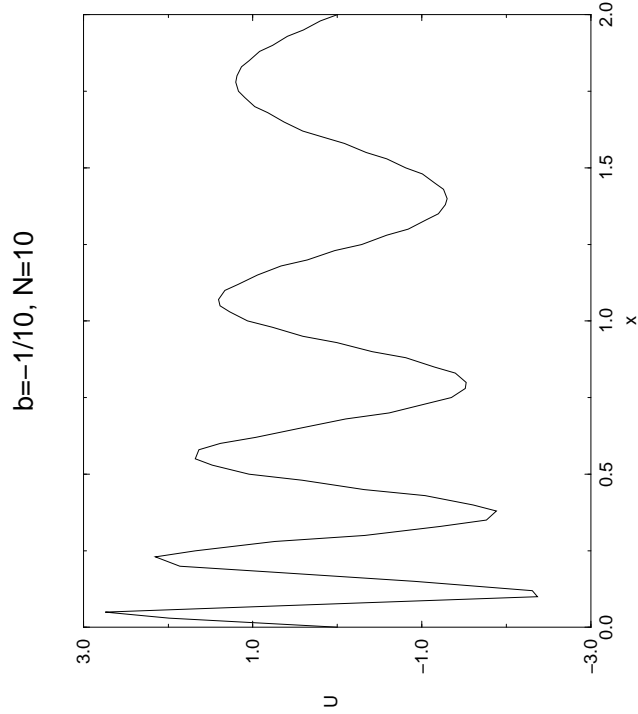}
\caption[]{Neutral mode in the WKB approximation in the limit $Fr = 
\Omega / N \to
0$ and $\beta\to\infty$ for $\beta Fr S/\Omega=-1/5$ and
$k_z=3\pi/2$.}
   \label{fig:figure2}
\end{figure}

Upon substitution of expression (\ref{TXa}) for
$\Theta_X$ in (\ref{exu0}), one obtains
\begin{equation}
{\partial \over \partial_X} \log\left \vert{U_0^2 H \Theta_X \over
(2i+\bar \sigma \Theta_X)}\right \vert \pm {2 i \over \beta k_z Fr}
\left[{k_z^2 Fr^2 \over \bar \sigma(1 - Fr^2 \beta^2 \bar \sigma^2)^{1/2}}+
{2(1 - Fr^2 \beta^2 \bar \sigma^2)^{1/2} \over \bar \sigma^3} .
\right]=0 \label{logu0}
\end{equation}
The identity
\begin{equation}
{2(1 - Fr^2 \beta^2 \bar \sigma^2)^{1/2} \over \bar \sigma^3}\equiv
{\partial \over
\partial_X} \left[{(1 - Fr^2 \beta^2 \bar \sigma^2)^{1/2} \over \bar
\sigma^2}\right]
-{\beta^2 Fr^2 \over \bar \sigma(1 - Fr^2 \beta^2 \bar \sigma^2)^{1/2}}
\end{equation}
is used to rearrange the terms between the brackets in Eq.
(\ref{logu0}). The only term which remains non-integrated is 
proportional to $(k_z^2-\beta^2)Fr^2$ and can be
neglected when the axial
and azimuthal wavenumbers have the same order of magnitude and $Fr
\to 0$. Under these simplifications
\begin{equation}
U_0= C {\bar \sigma^{1/2} \over (1 - Fr^2 \beta^2 \bar \sigma^2)^{1/4}}
\exp \left({\mp 2i \over \beta k_z Fr \bar \sigma^2}(1 - Fr^2 \beta^2
\bar \sigma^2)^{1/2} \right)  . \label{Ub}
\end{equation}
Under our approximations the exponential factor in (\ref{Ub}) can be
set equal to unity.
Introducing the notations $Q=1 - Fr^2 \beta^2 \bar \sigma^2$ and
$\gamma=k_z/\varepsilon \beta Fr$,
the velocity component U is expressed as
\begin{equation}
U={\bar \sigma^{1/2} \over Q^{1/4}} \left[ A \exp \left(i \gamma Q^{1/2}\right)
+B \exp \left( -i \gamma Q^{1/2}\right) \right] .
\end{equation}
The unknown coefficients $A$ and $B$ will be determined by
satisfying the boundary conditions
at $x=\pm 1$. Since in the limit $\beta \gg 1$ with $\beta Fr \sim
O(1)$, we have $P=Q^{1/2}U$
one can indifferently impose $U=0$ or $P=0$ at the boundaries, which
is of some importance in the
astrophysical context where it is usually considered as more
realistic to impose boundary conditions
on the pressure.
We shall define the a priori complex quantities
\begin{eqnarray}
Q_1 \equiv Q(x=1)=1 - \beta^2 Fr^2 (\bar \omega-\varepsilon)^2 \\
Q_2 \equiv Q(x=-1)=1 - \beta^2 Fr^2 (\bar \omega+\varepsilon)^2
\end{eqnarray}
with $\bar \omega=\bar \omega_r + i \bar \omega_i$. The algebraic
system for $A$ and $B$
resulting from the boundary conditions $U=0$ or $P=0$ has non-trivial
solutions only
if the following relation is satisfied
\begin{equation}
\sin \gamma (Q_1^{1/2}-Q_2^{1/2})=0
\end{equation}
which implies that its real and imaginary parts vanish
simultaneously. We were not able to find simple general expressions
for unstable modes, but we can exhibit analytical solutions
corresponding to the neutral modes ($\bar \omega_i=0)$
so that $Q_1$ and $Q_2$ are both real quantities. Moreover, we shall
focus on a particular
solution characterized by a fixed value of the frequency $\bar
\omega_r$ chosen so that
either $Q_1=0$ or $Q_2=0$. In the former case we have
\begin{equation}
\beta Fr \bar \omega_r =\beta Fr \varepsilon \pm 1 . \label{omgr}
\end{equation}
When $Q_1=0$, the condition $U(1)=0$ implies that $A+B=0$, and the
condition $U(-1)=0$ reduces to
\begin{equation}
\sin (\gamma Q_2^{1/2}) =0, \quad \mbox{or} \quad \gamma Q_2^{1/2}=N
\pi \label{Npi}
\end{equation}
where $N$ is an integer and $Q_2=-4 \varepsilon \bar \omega_r \beta^2
Fr^2$ has to
be positive for its square root to be real. Substituting the expression
(\ref{omgr}) in the condition (\ref{Npi}) gives the relationship
\begin{equation}
b \equiv \beta Fr \varepsilon = {\pm 4 \over \left(4 + (N\pi/k_z)^2\right)}
\end {equation}
between the combination $b=\beta Fr \varepsilon$ and the axial
   wavenumber $k_z$ for a solution $U$ to exist in the form
\begin{equation}
U={\bar \sigma^{1/2} \over Q^{1/4}} \sin [(k_z/b) Q^{1/2}]
\end{equation}
with $Q = -b (1-x)[2+b(1-x)]$ when $-1<b<0$. Such solutions have been
plotted in Figs. 1a
and 1b for $b=-1/5$ obtained with the value $N\pi/k_z=4$. Thus
varying the value of $N$ allows us
to change the value of $k_z=N(\pi/4)$ and as a consequence the number
of zeros of $U$. The Figs. \ref{fig:figure2} and \ref{fig:MS3345f3}
corresponds to $N=6$ and $N=10$ respectively, and as expected the
number of zeros is proportional
to $N$. The Figs. \ref{fig:figure4} and \ref{fig:MS3345f5} drawn for
$b=-1/10$ obtained with the value
$N\pi/k_z=6$,  correspond respectively to $N=6$ and $N=10$.
It must be noticed that when the instability occurs via a Hopf
bifurcation ($\omega_r \ne 0$)
the eigenfunctions break the symmetry about $x=0$,
the amplitude being larger near one side of the interval, here
$x=-1$. This has already been
discussed by Knobloch (1996) who showed that system with O(2) symmetry can
exhibit either a steady state bifurcation or a  Hopf bifurcation.

\begin{figure}[hhh]
\centering
\includegraphics[width=8.5cm]{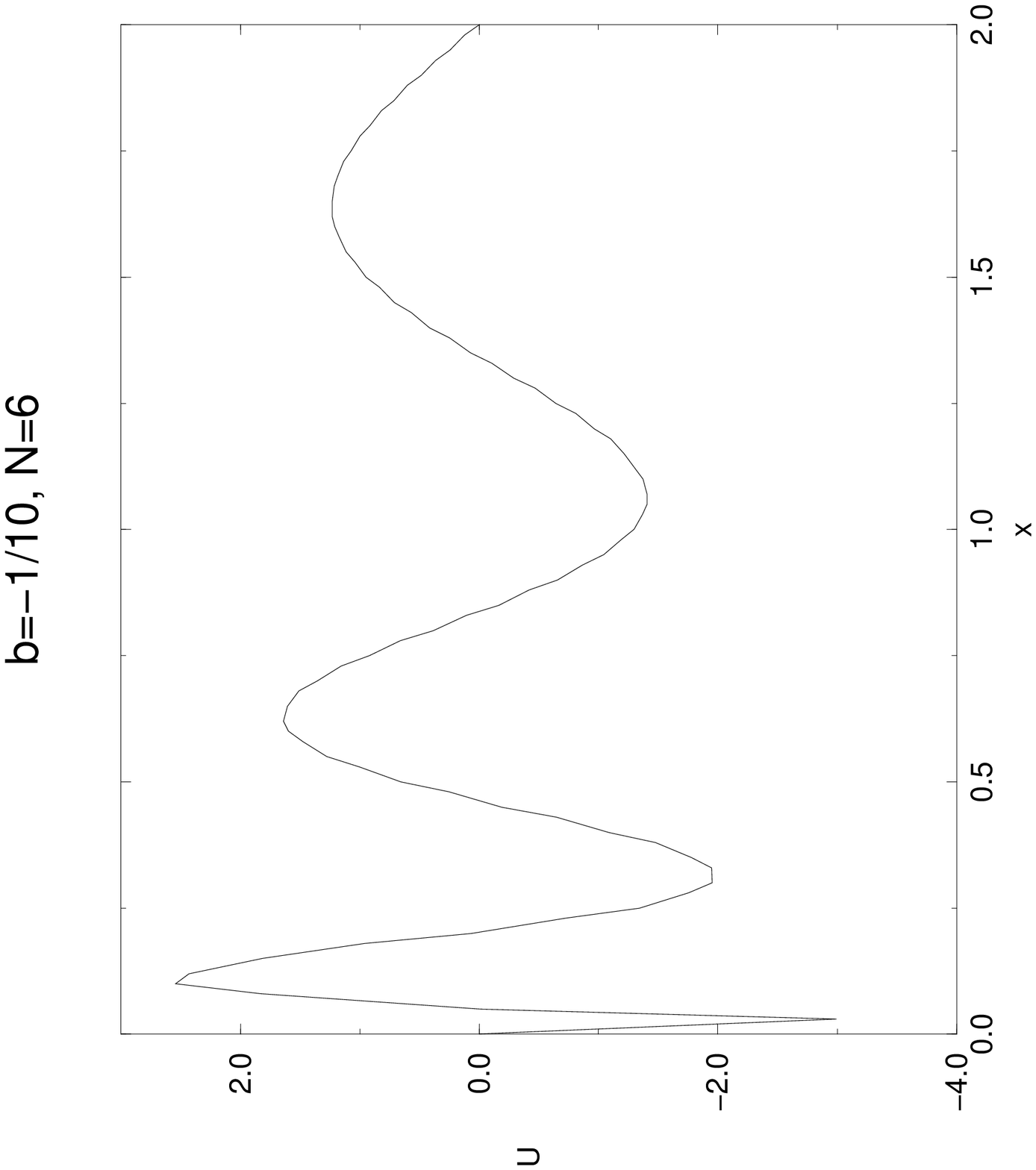}
\caption[]{Neutral mode in the WKB approximation in the limit $Fr\to
0$ and $\beta\to\infty$ for $\beta Fr S/\Omega=-1/5$ and
$k_z=5\pi/2$.}
   \label{fig:figure3}
\end{figure}

\begin{figure}[hhh]
\centering
\includegraphics[width=8.5cm]{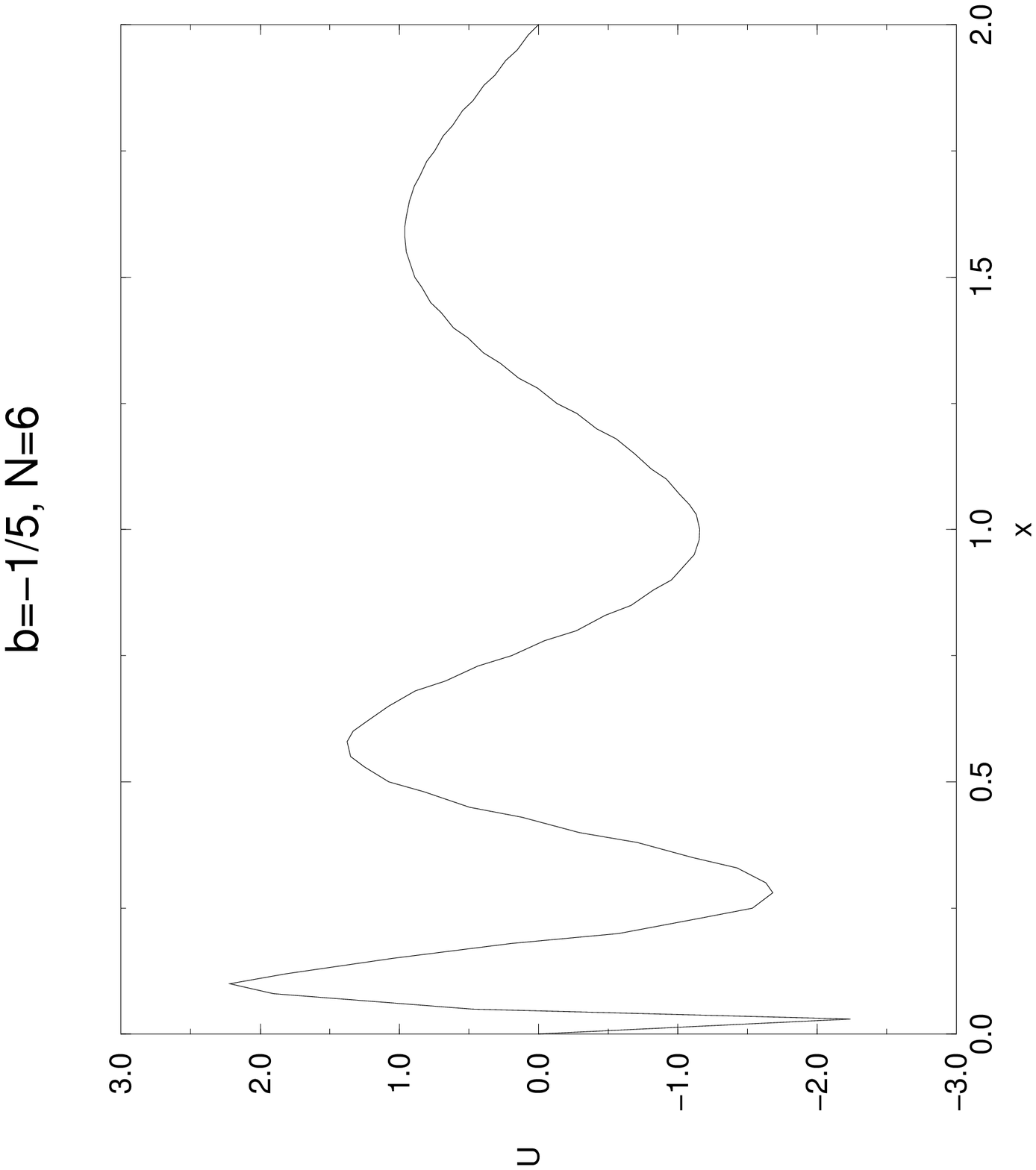}
\caption[]{Neutral mode in the WKB approximation in the limit $Fr\to
0$ and $\beta\to\infty$ for $\beta Fr S/\Omega=-1/10$ and $k_z=\pi$.}
   \label{fig:figure4}
\end{figure}

\begin{figure}[hhh]
\centering
\includegraphics[width=8.5cm]{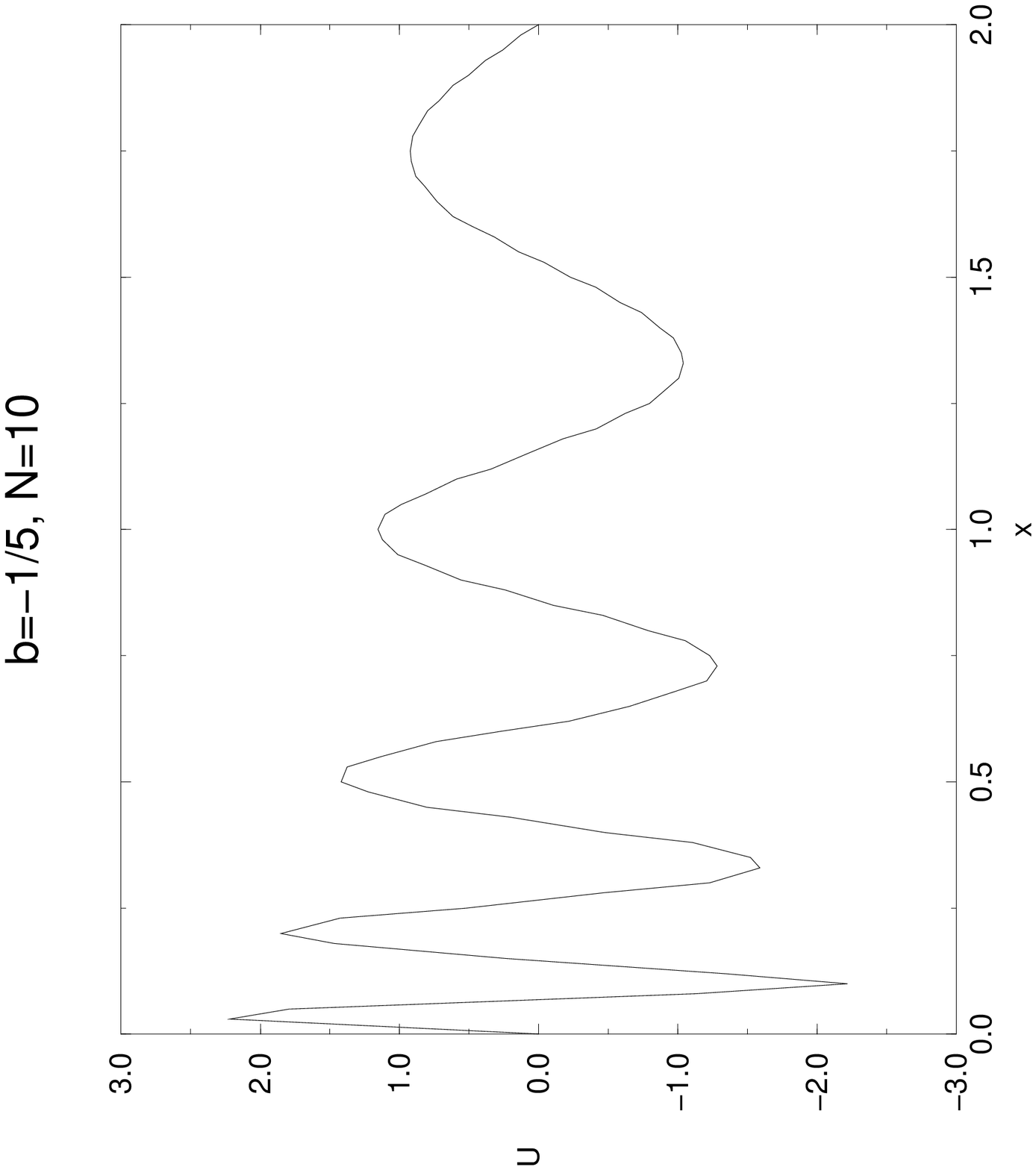}
\caption[]{Neutral mode in the WKB approximation in the limit $Fr\to
0$ and $\beta\to\infty$ for $\beta Fr S/\Omega=-1/10$ and
$k_z=5\pi/3$.}
   \label{fig:figure5}
\end{figure}

\section{Numerical study}

\subsection{Method}
The condition for stability derived in Section 2 is
valid for a wide class of boundary conditions, but only in the inviscid
limit.
The
influence of viscosity has been studied numerically by Molemaker et al. (2001)
and by Yavneh et al. (2001), in the limit $Pr\to\infty$. They
found that a small viscosity essentially does not change the results,
and only introduces a critical Reynolds number, above which the flow is
unstable. For example, at $\Omega/N=0.01$, $S/2\Omega=-2/3$, the
critical Reynolds
number is of order 4800, for rigid boundary conditions. In a
Keplerian disk, $S/2\Omega=-3/4$ and
$Pr$ is finite, so that
the results of Yavneh et al. do not apply. We thus performed a
numerical study
of equation (\ref{basicequationlocal}) in the plane case to
investigate the range of parameters
of astrophysical interest. The stability analysis was set into a
classical eigenvalue problem via Fourier transform in the $y$ and $z$
direction, and  discretization of the equation over collocation
points of the Chebyshev polynomials in the x-direction, over a domain
$-d/2\le x\le d/2$.
    The precision
of the solution is governed by the number of collocation points used in
the computation. We used typically 25 to 50 collocation points to get
satisfactory precision: doubling the resolution did not alter the
results. Spurious
normal modes were eliminated via interpolation, then cross-checked,  at
double resolution. In the following, we present results obtained using
stress-free boundary conditions in $y$ and $z$ (the case of rigid
boundary conditions is discussed in Yavneh et al. (2001).)\

Dimensional analysis of equation (\ref{basicequationlocal})
    shows
that the stability problem is controlled by four non-dimensional
parameters: the rotation number $Ro=2\Omega/S$, the Ekman number
$Ek=\nu/2\Omega d^2$ , the Froude number
$Fr=\Omega/N$ and the Prandtl number $Pr$. From these numbers, one can
also build another non-dimensional number of interest, the Reynolds
number, defined as:
$Re=\vert S\vert d^2/\nu=Ek^{-1} \vert Ro^{-1}\vert$. Using
the gap $d$ (equivalent to the typical scale of our perturbation, in
the disk case, see Section 2)
and $1/2\Omega$ as unit of length and time, we may  then write
(\ref{basicequationlocal}) as:
\EQA
-i\sigma_\ast u- v&=&-Dp+Ek\left(\partial_x^2-k^2\right)u\nonumber\\
-i\sigma_\ast v+(1+1/Ro) u&=&-i\beta p+
Ek\left(\partial_x^2-k^2\right)v\nonumber\\
-i\sigma_\ast w +\frac{Fr^{-2}}{4} \theta&=& -i k_zp
+Ek\left(\partial_x^2-k^2\right)w,\nonumber\\
-i\sigma_\ast \theta&=&
w+\frac{Ek}{Pr}\left(\partial_x^2-k^2\right)\theta,\nonumber\\
Du+i\beta v+i k_zw&=&0,
\label{basicequationlocaladim}
\ENA
where $\sigma_\ast=\sigma/2\Omega$.

    Our numerical study
was focused on the determination of maximal growth rate and
wavenumber, for a given value of the four parameters. Due to the
large number of independent parameters, it is difficult to completely explore
  the instability phase space. Moreover, we found out that
for $Ro<-2$, the search for maximal growth rate
was increasingly time consuming due to the exponential narrowing
of the instability branch around the maximal value. This phenomenon
had been described in Molemaker et al (2001). We thus mainly focused
our analysis on the case $Ro>-2$, with special emphasis on the case
$Ro=-4/3$ corresponding to the Keplerian case.

\subsection{Reminder}

A few results regarding special values of the parameters can already
be drawn from previous studies: in the unstratified case
$Fr\to\infty$, Lezius and Johnson
(1971) found that the neutral stability curve is given by
\EQ
Re^2 Ro (Ro+1)=-1706,\quad Fr=\infty.
\label{lezius}
\EN
In the inviscid case $Ek=0$, Molemaker et al. (2001) found that for
$Fr<1$, the maximal growth rate and vertical wavenumber scale as:
\EQA
\frac{\omega_i}{2\Omega}&=&-2 Ro^{-1} \beta d\exp(2Ro),\nonumber\\
k_z d&=&\frac{1}{2 Fr}\frac{Ro}{\sqrt{Ro(Ro+1)}}.
\label{molemaker1}
\ENA
In the case $Fr>1, Ek=0$ they found that both $k_zd$ and
$\omega_i/2\Omega$ decay linearly with $1/Fr$.

\subsection{Influence of viscosity}
\begin{figure}[hhh]
\centering
\includegraphics[width=8.5cm]{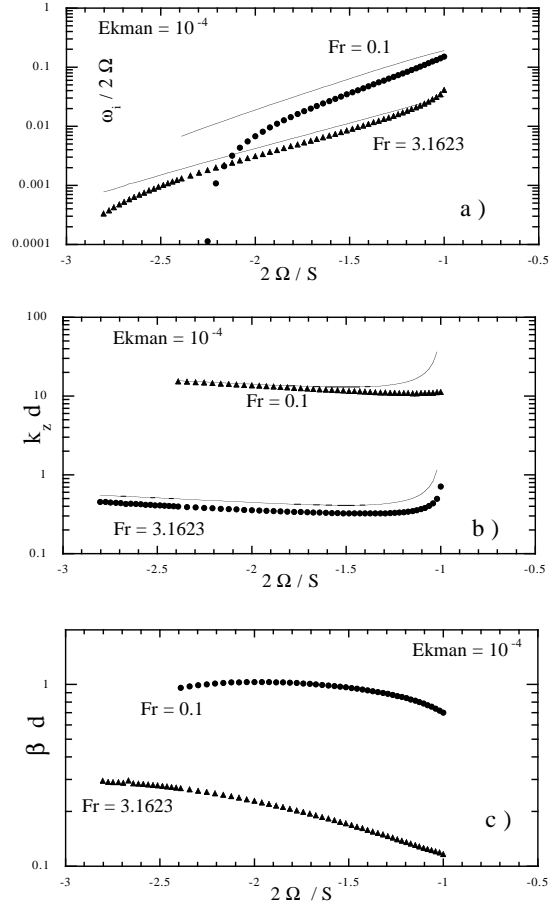}
\caption[]{Maximal growth rate (2a), corresponding vertical wavenumber (2b) and
azimuthal wavenumber (2c) as a function of the rotation number $2 \Omega/S$ at
$Ek=10^{-4}$, $Pr=\infty$ and for different stratifications. The line is
the result of Molemaker et al. (2001), obtained at $Ek=0$.}
   \label{fig:figure6}
\end{figure}

\begin{figure}[hhh]
\centering
\includegraphics[width=8.5cm]{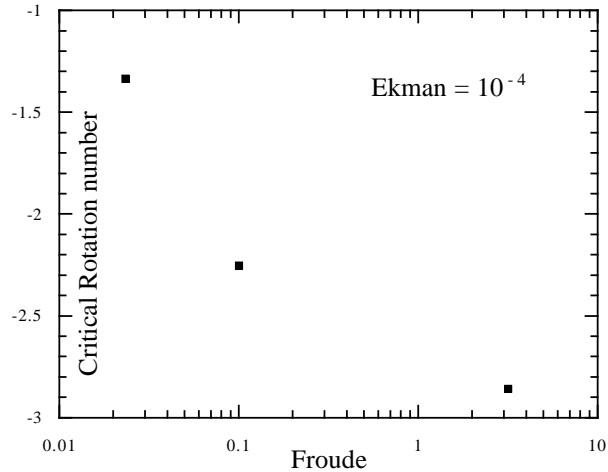}
\caption[]{Critical rotation
number vs. Froude number $(\Omega / N)$,
at $Ek=10^{-4}$.}
   \label{fig:figure7}
\end{figure}

To study the influence of viscosity on the instability, we set
$Pr=\infty$ and vary the Ekman number, for different rotation number
and stratifications. In Figure \ref{fig:figure6}, we show the
non-dimensional
optimal growth rate, vertical and azimuthal wavenumber as a function
of the rotation number, at $Ek=10^{-4}$, for two different
stratifications $Fr=3.1623$ and $Fr=0.1$. The inviscid result
(\ref{molemaker1}) is also displayed for comparison. One sees that
viscosity tends to decrease optimal growth rate and wavenumbers.
Moreover, it induces a finite critical value of the rotation number
below which the optimal growth rate becomes negative. This critical
rotation number depends on the stratification
(Figure \ref{fig:figure7}). It decreases from $-1.3$ at
$Fr=0.02$ up to $-2.85$ at
$Fr=3.1623$.
We note on the figure a tendency towards leveling at the value
$Ro=-3$, which may indicate a critical rotation number of $-3$ for
$Ek=10^{-4}$, below which there is no instability, whatever the
stratification.\

\begin{figure}[hhh]
\centering
\includegraphics[width=8.5cm]{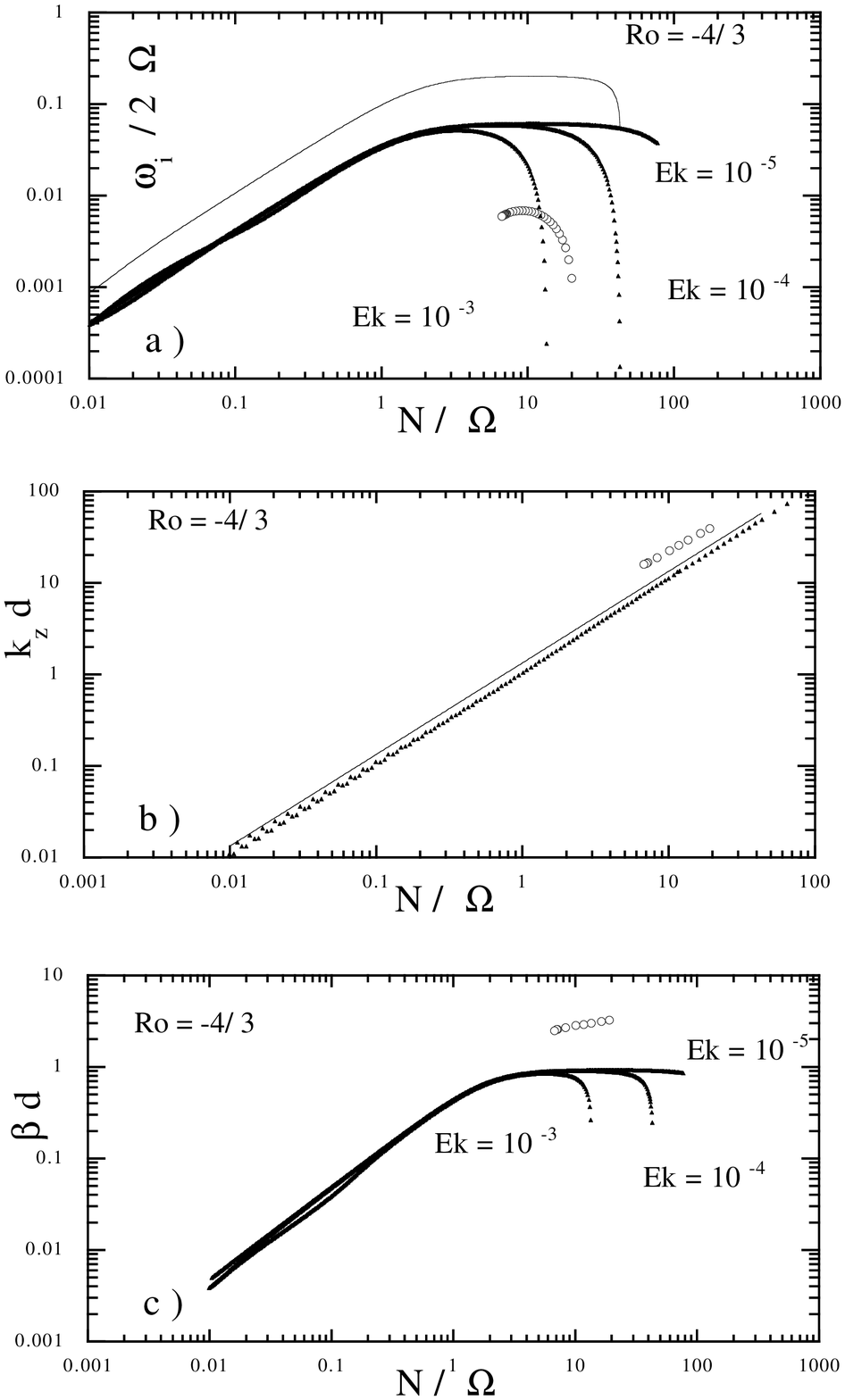}
\caption[]{Maximal growth rate (4a), corresponding vertical wavenumber (4b) and
azimuthal wavenumber (4c) as a function of the inverse Froude number at
$Ro=-4/3$, $Pr=\infty$ and for different viscosity. The filled symbols
correspond to the optimal eigen mode; the open symbols correspond to a
member of the ``large $\beta$ family", computed at $Ek=10^{-5}$. The
lines refer to the theoretical formulae proposed by  Molemaker et al. (2001)
for the optimal mode: $\omega_i/ 2\Omega= \beta d e^{2 Ro}$ and $k_z
d=Ro/2 Fr \sqrt{1+Ro}$.}
   \label{fig:figure8}
\end{figure}

\begin{figure}[hhh]
\centering
\includegraphics[width=8.5cm]{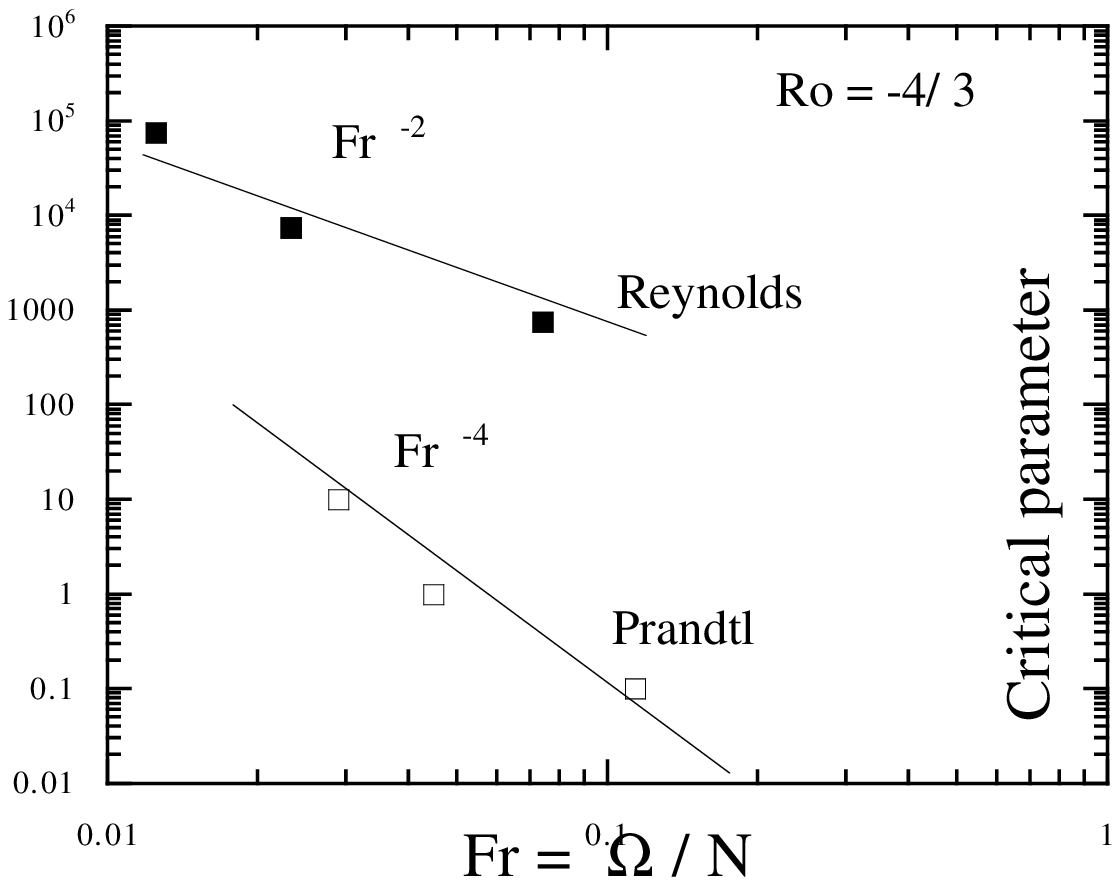}
\caption[]{Variation of Reynolds and Prandtl numbers on the
neutral stability lines at $Ro=-4/3$ as a function of the Froude
number.}
   \label{fig:figure9}
\end{figure}

In Fig \ref{fig:figure8}, we show the optimal growth rate
and wavenumbers as
functions of stratification, at $Ro=-4/3$, for different values of
$Ek$. For comparison, the results obtained by Molemaker et al (2001)
at $Ro=-3/2$ and $Ek=0$ (for different boundary condition!) are also
displayed. The growth rate tends to be proportional to $N$ for large
Froude numbers, but becomes a fraction of the dynamical frequency $\Omega$
for $Fr>1$. One sees that the viscosity (and boundary conditions) has
little influence on the
vertical wavenumber, which scales as $1/Fr$, while it introduces a
sharp cut-off at small
Froude numbers in both $\omega_i$ and $\beta$. This cut-off is pushed
towards larger stratification with increasing viscosity. The cut-off
defines a viscosity-dependent critical Froude number, below which no
instability is present. Its dependence on the Reynolds number is
approximately a power law
$Fr_c\sim Re^{-1/2}$, as shown in Fig. \ref{fig:figure9}.\

The variation with stratification indicates that at small
stratification, the growth rate is proportional to $N$, while at
large stratification, it is proportional to $\Omega$, i.e.
independent of the stratification.

\subsection{Influence of Prandtl number}
\begin{figure}[hhh]
\centering
\includegraphics[width=8.5cm]{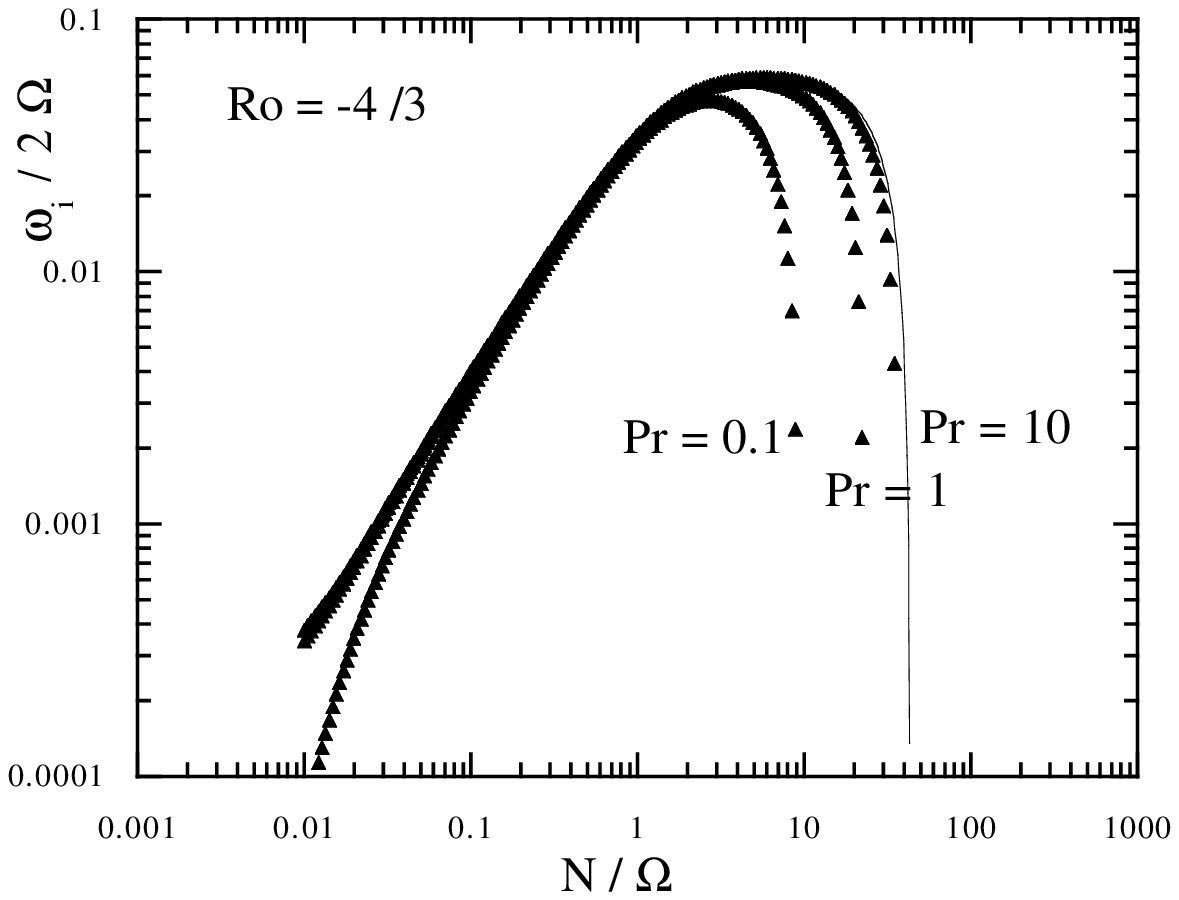}
\caption[]{Maximal growth rate as a function of the inverse Froude number
at $Ro=-4/3$, $Ek=10^{-4}$ and for different Prandtl numbers. The line
is the result at $Pr=\infty$, drawn for comparison.}
   \label{fig:figure10}
\end{figure}

For finite Prandtl number, the stratification agent
(for example the temperature) undergoes a
diffusive process, which can have a stabilizing influence.  To study
this effect, we fix $Ro=-4/3$ and $Ek=10^{-4}$, and decrease the
Prandtl number from $Pr=\infty$ to $Pr=0.1$. The resulting optimal
growth rate and wavenumbers are shown in Fig.
\ref{fig:figure10}. One sees that
decreasing the Prandtl number at fixed Ekman number provides qualitatively
the same effects as when increasing the Ekman number (previous
Section): as $Pr$ is decreased, the critical Froude number is shifted
towards higher values. We have checked that this similarity of
behavior also extends to optimal wavenumbers: no incidence on
vertical wavenumber, varying cut-off for $\beta$. The variation of the
critical Froude number with Prandtl number is shown in Fig.
\ref{fig:figure9}. The variation
of $Fr_c$ with $Pr$ is less steep than for $Re$, namely $Fr_c\sim
Pr^{-1/4}$.\

\begin{figure}[hhh]
\centering
\includegraphics[width=8.5cm]{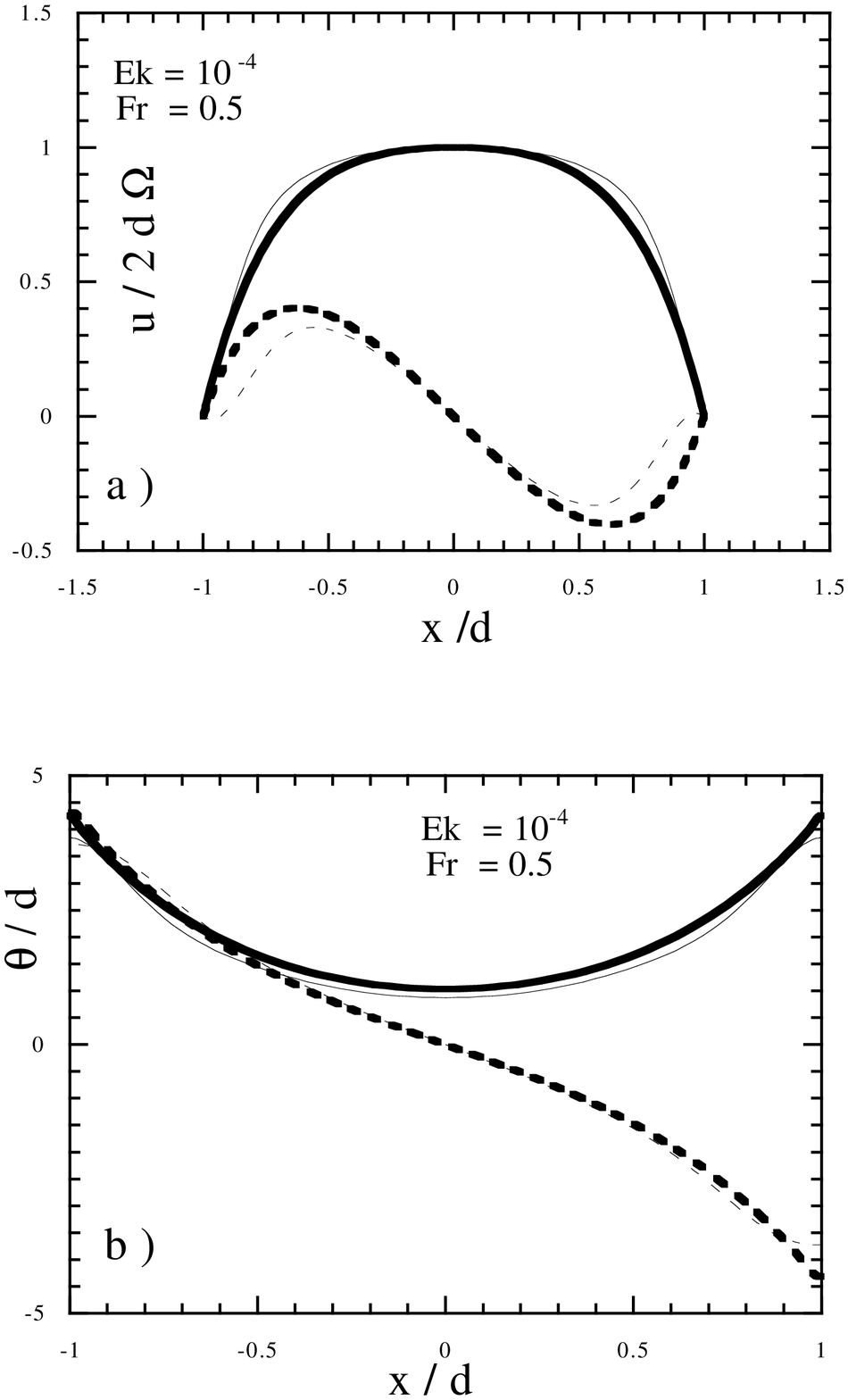}
\caption[]{Real (continuous line) and imaginary (dotted line) part of
the optimal eigenfunction for the radial velocity $u$ (Fig 7a) and
for the scaled entropy fluctuation $\theta$ (Fig 7b),
as a function of the $x$-coordinate, at $Ro=-4/3$, $Fr=0.5$,
$Ek=10^{-4}$, and for $Pr = 0.1$ (thin line) and 10 (thick
line).}
   \label{fig:figure11}
\end{figure}

For illustration we also display in Fig. \ref{fig:figure11} a typical
eigenmode solution for the radial velocity and entropy perturbation
for $Ro=-4/3$, $Ek=10^{-4}$, $Fr=0.5$ and two values of the Prandtl
number. One sees that these eigenfunctions vary smoothly over the
channel width and that
the change with Prandtl number is rather smooth and unimportant.

\subsection{Influence of azimuthal wavenumber}

\begin{figure}[hhh]
\centering
\includegraphics[width=8.5cm]{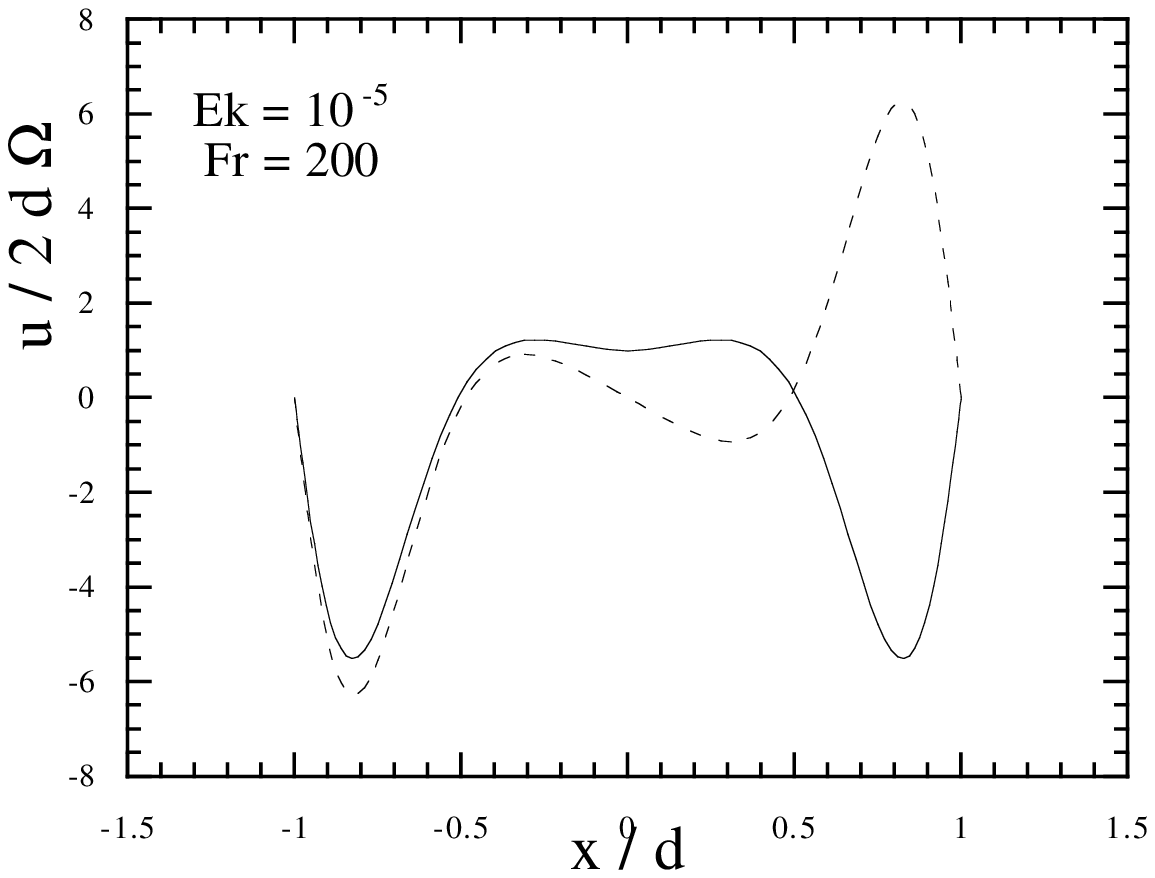}
\caption[]{Real (continuous line) and imaginary (dotted line) part of
a "large $\beta$"  eigenfunction for the radial velocity $u$,
as a function of the $x$-coordinate, at $Ro=-4/3$ $Fr=200$
$Ek=10^{-5}$, $\beta d=3.3$, $k_z d=41$, $\omega_i/2\Omega=0.0013$
and for infinite Prandtl number.}
   \label{fig:figure12}
\end{figure}

Up to now, we have focused only on optimal modes (those with highest
growthrate). These modes are characterized by a low azimuthal
wavenumber. In the application to astrophysical disks, it may be
interesting to look for modes with larger azimuthal wavenumber, which
may be more realistic for very thin disks (for which $\beta d\sim
r/H\gg 1$). Finding such modes is very time consuming using our
numerical procedure (which is optimized for the detection of optimal
growth rate). So, we focused on the case $Ro=-4/3$ and only explored
the case $Ek=10^{-5}$, $Pr=\infty$. In that case, we were able to
capture one member of the large $\beta$ family whose characteristics
are shown in Fig. \ref{fig:figure8} for easier
comparison with the optimal mode. The azimuthal and vertical wavenumber
of this mode are roughly twice those of the optimal
family. Its growth rate is about 10 times smaller. The corresponding
velocity profile shows a double oscillation within the channel,
see Fig. \ref{fig:figure12}. The mode seems to exist only at
rather large stratification $Fr>1$. All these features are reminiscent
of the ``large $\beta$ mode" found in the Taylor-Couette configuration
by Molemaker et al. (2001). These modes are characterized by a
``radial" wave number $k>2$, thereby producing $k$ oscillations of the
eigenmode in the radial direction. The wavenumber roughly follows
$k^2=\beta ^2 k_z^2$. Their growth rate, in the inviscid limit, scales
as $\vert \Omega\vert k^{-1/4} e^{3 Ro}$. With $k=2$, this gives
$\omega_i/2\Omega=0.008$ in rather good agreement with our finding.
This strongly suggests that these ``large $\beta$ -modes" also exist in
the plane Couette case. Since their growthrate is still a significant
fraction of the rotation period, these modes are probably also quite
relevant for astrophysical disks.

\subsection{Comparison with experiments}

\begin{figure}[hhh]
\centering
\includegraphics[width=8.5cm]{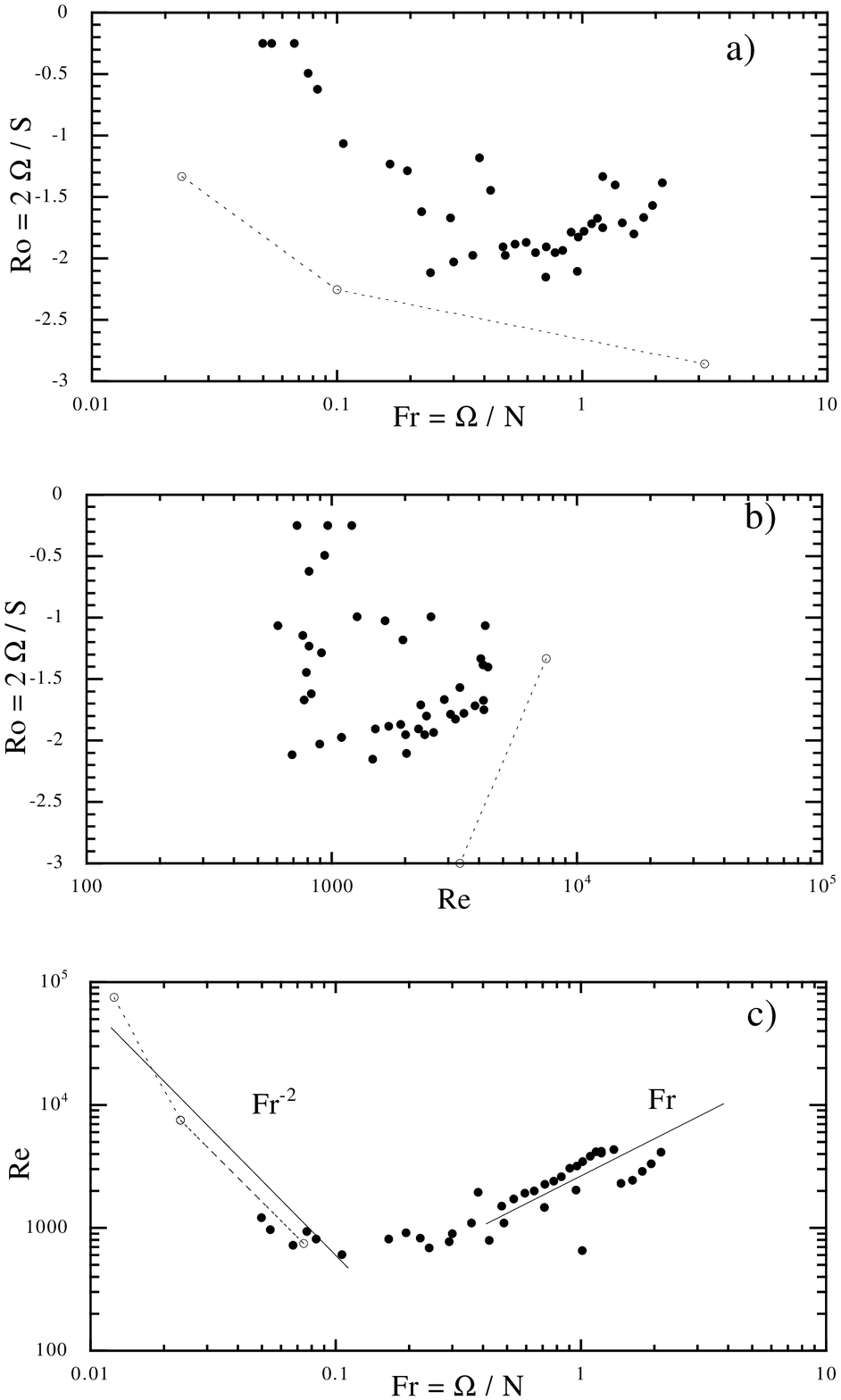}
\caption[]{Neutral-stability curves ($\omega_i=0$) in the stratified
Taylor-Couette flow
experiment of Boubnov and Hopfinger (1995) in the small gap limit, in
parameter space; a): in the plane $Ro,Fr$; b) in the plane
$Ro,Re$; c) in the plane $Re, Fr$.
The filled circles are the experimental data. The open circles are
the numerical data obtained in the present paper.}
   \label{fig:figure13}
\end{figure}

It is interesting to compare our results with experimental data
obtained in the same conditions (small gap limit) by Boubnov and
Hopfinger (1995). We note that in this experiment the
boundary conditions are probably rigid (in contrast
with our numerical analysis). This comparison can then be used to
probe the influence of boundary conditions. A difficulty arises because the
experiments only provide
data on the critical line, on which $Ro$, $Fr$ and $Re$ vary
simultaneously. We thus analyze the three possible planes, in Fig.
\ref{fig:figure13}. In panel a), we observe a confirmation of
the stabilizing role played by stratification: weak stratification
tends to increase the range of instability, while large
stratification (small Froude number) tends to restrict the domain of
instability. The experimental points do not overlap with our numerical points, since the latter are performed at constant, and slightly sma
ller Ekman number. In panel b), we do not observe any clear
dependence of the rotation number on Reynolds number (different
instability branches may be present). At $Ro=-4/3$, the Keplerian
value, the critical Reynolds number seems to be between  $4000$ and
$10^4$, well below typical Reynolds number of e.g. circumstellar
disks (Hersant et al, 2003). In panel c) one sees the dependence of
the critical Reynolds number as a function of the stratification. At
small Froude numbers, the experimental data seem to confirm the
$Fr^{-2}$ scaling obtained numerically. At large Froude number, we
observe another, linear scaling, of the Reynolds number with the
Froude number, $Re_c\sim 2000 Fr$.
This number will
be used below in our discussion.

\section{Application to astrophysics}
\subsection{Stratification in disks}
In previous Sections, we have shown that all flows with $S/2\Omega < 0$
are linearly unstable with respect to non-axisymmetric perturbations
in the presence of vertical stable stratification.

To decide whether an astrophysical disk is stably stratified, on has 
to solve for the vertical structure of the disk, which itself depends 
on how turbulent kinetic energy is transformed into heat through 
viscous friction. Most prescriptions for this heat release lead to 
stable stratifications.

One exception may be the central part of disks around black holes, 
where radiation pressure exceeds gas pressure, and where for this 
reason entropy decreases with distance from the equatorial plane, 
triggering  thermal convection (Bisnovatyi-Kogan \& Blinnikov 1977). 
But this does not take into account the illumination of the disk by 
hard X-rays, which is observed.
The presence of thermal convection has also been suggested in disks 
around young stellar objects
(Lin \& Papaloizou 1980). However, here again a natural process leading to
stable vertical stratification is the disk illumination by the
central object (Chiang \&  Goldreich 1997).

A rough estimate of the Brunt-V\"ais\"al\"a frequency in stably 
stratified disks may be obtained in the thin disk approximation:
\begin{equation}
N^2 ={g \over H_P}  \left[\nabla_{\rm ad} - \nabla \right] =
\Omega ^2  \left({z \over H}\right)^2 \left[\nabla_{\rm ad} - \nabla \right]
\end{equation}
where the actual temperature gradient $d \ln T /  d\ln P$ is 
determined by the opacity law;  the adiabatic gradient $\nabla_{\rm 
ad}= (\partial \ln T / \partial \ln  P)_{\rm ad}$ takes the value 0.4 
for perfect atomic or ionized gas. Typically $[\nabla_{\rm ad} - 
\nabla] \approx 0.1$, and thus $N / \Omega \sim 0.3$.
This value is confirmed by detailed calculations.

Radiative numerical simulations of protostellar disk (D'Alessio et 
al. 1998) reveal that at $r=0.5$
a.u. the temperature is of the order of $1000K$ at the disk scale height
and $300K$ in the midplane, leading to $N/\Omega\sim 0.3$. In this
regime, the critical Reynolds number is of the order of $1000$, the
growthrate is about one percent of the rotation frequency and the
typical vertical and azimuthal wavenumbers are about $0.1/d$. The
largest allowed $d$ in this case is $0.1 H$ leading to vertical scale
and azimuthal wavenumber of the order of unity. The typical instability
should then take the form of a spiral mode.

We can thus conclude that, in astrophysical disks, stable 
stratification is the rule rather than the exception.
If this were not the case, one would invoke thermal convection as the 
obvious cause for the turbulence which ensures the accretion on the 
central object.
\subsection{Magnetic field vs. stratification}

Our findings are reminiscent of what occurs in the presence of an
axial magnetic field. It is therefore interesting to examine the
degree of similarity between the two instability mechanisms using the
same tools. For this, we recall the results obtained by Chandrasekhar
(1960). First, unlike the hydrodynamical instability in stratified
disks,
the magneto-rotational instability has axisymmetric unstable modes.
For these modes, the sufficient stability condition  is:
\EQ
S/2\Omega>-\frac{1}{4}\frac{I_1}{k_z^2I_0}
\frac{\Omega_A^2}{\Omega^2},\quad\rm{STABILITY}
\label{analogcondi}
\EN
where $\Omega_A=\sqrt{k_z^2 \mu B_z^2/4\pi\rho}$ is the Alfv\`en
frequency. Comparing this condition with (\ref{condiinstability}), we
see strong similarities with
the stratified case, the Brunt-V\"aiss\"al\"a frequency being just
replaced by the Alfv\`en frequency.
For example, strong magnetic fields tend to inhibit instability, in the
same way as strong stratification, whereas in the limit of the weak
magnetic field $S/2\Omega>0$ is a condition for stability,
as  for stratified, rotating shear flows.
Keplerian flows, having $S/2\Omega=-3/4$ do not satisfy this
stability criterion, and are therefore liable
to both type of instabilities.\

In a recent detailed numerical study of the magneto-rotational
instability, Willis and Barenghi (2002) found that the critical
Reynolds number at infinite Prandtl number is of the order of 20, that is
one order of magnitude less than for the strato-rotational
instability discussed here. They also noted that the variation of the
critical Reynolds number with magnetic Prandtl number is
$Re_c\sim 100/\sqrt{P_m}$, similar to the law found in Fig. 5 for
stratification. However, in a typical disk, the magnetic Prandtl
number is much lower than the Prandtl number. For example, in a
protoplanetary disk, it is less than $10^{-5}$ (R\"udiger \& Zhang
2001). This means that a typical critical Reynolds number for the
magneto-rotational instability will be greater than $30000$, hence
larger than the critical Reynolds number for strato-rotational
instability.

\subsection{Barotropic vs baroclinic instability?}

In the present paper, we only focused on perturbations with
typical radial scale small compared to the vertical scale height,
leading to a barotropic description. An opposite limit would be to
consider perturbations with small vertical scale, leading to a
situation where only the vertical dependence is considered so that
$\Omega$ varies
with the axial coordinate $z$. In this situation, one may expect new
instabilities to arise, due to the vertical shear. More generally,
when the rotation departs from cylindrical, it may induce
an axisymmetric baroclinic instability which has been described
earlier in stellar interiors by Goldreich and Schubert (1967) and
Fricke (1968). The application of this instability to disks has been
recently done by Urpin (2003), who showed that the instability
is linear, and that it proceeds on a thermal time-scale. More generally,
baroclinic instabilities occur when there is an inclination between
the density and pressure term, breaking the vorticity conservation.
As discussed by Klahr and Bodenheimer (2003), this effect is likely
to occur in any realistic disk. This means that in general,
astrophysical disks are subject to both barotropic (as shown in this
paper) and baroclinic instabilities. The interplay between these two
instabilities is a fascinating subject, and has been widely studied
in the oceanographic context. For upwelling flows it was found for
example that barotropic instabilities are important in the early
stages of the dynamics, while baroclinic modes dominate the late
stages. The growthrate and wavenumbers are in between the growthrate
and wavenumber of each instability (Tadepalli \& Fertziger 1998).
Clearly, a similar study would be welcome in the astrophysical
context.

\section{Conclusion}
In this paper, we have used analytical and numerical methods to study
the instability of Keplerian-like flow in the presence of a stable
vertical stratification,  for perturbations obeying rigid, 
stress-free or periodic boundary conditions in the radial coordinates 
that lie less than one scale height apart so as to justify an 
incompressibility approximation.
This instability is non-axisymmetric. Its growth-rate is a fraction of
the rotation rate at small stratification.
This `strato-rotational' instability is purely hydrodynamical
and does not require the presence of any magnetic field,
whatever small. Therefore, it is especially relevant in the context
of weakly ionized disks, such as the primitive solar nebula.
It is of course also relevant in other Keplerian disks, since the
critical Reynolds
number to trigger this instability is of the order of $10^3$, which is
less than the critical Reynolds number for both the magneto-rotational
instability and the finite-amplitude hydrodynamic instability. It may
also interplay with baroclinic instabilities, resulting in mixed
instabilities similar to those observed in oceans.
However the question of which instability occurs first has
little relevance, partly because we do not know the initial
conditions, but mainly because there is no reason why
the instability which has the fastest linear growth will
dominate in the fully nonlinear state.
What one should rather ask is what kind of turbulence exists
at the high Reynolds numbers which characterize astrophysical disks.
A partial answer to this question may already be found from
Taylor-Couette laboratory measurements (Dubrulle et al. 2004), which
provide information about the hydrodynamical regime of rotating
shear flow, as well as the influence of magnetic field or
stratification on the transport properties. When applied to
circumstellar disks, this provides parameter free-predictions about
accretion rates and fluctuations which are in good agreement with
observations (Hersant et al. 2003). For the influence of processes
more specific to astrophysics (radiative transport, etc.) one will
probably have to wait until the numerical simulations are able to
reach Reynolds numbers of order $10^6$ or more.

\begin{acknowledgements}
We thank F. Daviaud and O. Dauchot for many suggestions and
discussions, M. Tagger, A. Brandenburg, S. Balbus and J. Goodman
for comments on an earlier version of the manuscript, and P.-Y.
Longaretti and an anonymous referee for references about sheared modes.
\end{acknowledgements}

\end{document}